**Effects of Virtual Hand Representation on the Typing Performance, Upper Extremity Angle, and Neck Muscle Activity during Virtual Reality Typing**

Mobasshira Zaman[1] and Jaejin Hwang[1*]

[1]Department of Industrial and Systems Engineering, Northern Illinois University, DeKalb, IL, USA

*Corresponding author:

Jaejin Hwang
Department of Industrial and Systems Engineering
Northern Illinois University
590 Garden Road, EB230
Tel: +1-815-753-9980
E-mail: jhwang3@niu.edu

**Effects of Virtual Hand Representation on the Typing Performance, Upper Extremity Angle, and Neck Muscle Activity during Virtual Reality Typing**

**Abstract**

This study evaluated the effect of virtual hand representation on the typing performance, upper extremity angle, neck muscle activity, and usability during virtual reality (VR) typing. A total of 15 participants (7 females and 8 males) performed a typing task with and without virtual hand representations. The optical motion capture data was utilized to capture the upper extremity angles, and electromyography device was used to collect the neck muscle activities (left and right splenius capitis). The results showed that the typing performance, upper extremity angle, neck muscle activity, and usability were significantly different with and without virtual hand representations. With the virtual hand representation, net typing speed (WPM) and usability increased significantly by 171.4% and 25% compared to the without hand representation. Without the virtual hand representation, participants showed increased wrist extension, and decreased right shoulder abduction angles. The variability of the neck rotation was increased while typing without the virtual hand representation. The neck muscle activities were increased with the virtual hand representation. The results suggest that typing with the virtual hand representation could positively affect the typing performance and usability and reduce the risk of the musculoskeletal disorders of the upper extremity. Future study could explore the effect of the virtual hand representation for users with varying levels of typing skills.

**Effects of Virtual Hand Representation on the Typing Performance, Upper Extremity Angle, and Neck Muscle Activity during Virtual Reality Typing**

## 1. Introduction

Virtual reality (VR) has become an advanced technology that caters to the future requirements of businesses in terms of design, training, and corporate communication. VR allows users to mentally experience flying, swimming, running, and walking through different structures (Whitman et al., 2004). VR presents numerous potential applications in various sectors such as entertainment, government, education, and virtually every other industry. (Patel and Cardinali, 1994). In the past decade, the reason behind VR gaining popularity as a valuable tool in a lot of fields has been because of its unique and immersive experience in our daily lives that has the potential to change the way we interact with one another, learn, and work (Grubert et al., 2018a). Furthermore, VR has been gaining popularity in recent years because of the development of the natural interface (Poth, 2021).

Typing is one of the most common tasks in the workplace and if VR is being used in the workplace comfortable typing condition needs to be provided for work efficiency. Although the use of a conventional keyboard for VR for typing tasks can provide a number of advantages, such as increased familiarity, reduced training, and improved accuracy it also poses several challenges and concerns (Grubert et al., 2018b; Knierim et al., 2018). Because not only the advanced apps and interface but also the immersed three-dimensional virtual physical appearance was a pivotal factor in satisfying the users with the realistic interface (Bowman et al., 2002; Shin et al., 2013). The virtual hand representation can have a significant impact on the user's typing performance, posture, and overall user experience (Zaman, 2022). Therefore, it is important to study the effects of virtual hand representation on various aspects of typing tasks.

In typing the higher wrist, angular deviation at joints can cause compression on the wrist and reduce the blood flow, which can increase the risk of carpal tunnel syndrome (Rempel et al., 2008). Shoulder angle can be linked to the wrist postures, i.e., shoulder abduction angle can minimize the ulnar deviation of the wrist, which reduces awkward posture. Moreover, maintaining an awkward shoulder posture while typing and repetitive movement can lead to tendinitis, bursitis, rotator cuff injuries, frozen shoulder, and thoracic outlet syndrome (Nag, 2019). According to Huang et al. (2012), increased muscle activity in the neck and shoulder regions was positively related to typing speed as well as accuracy when using a conventional keyboard (Huang et al., 2012).

This study aimed to identify the reduction in physical stress on the shoulder, and wrist, evaluate the muscle activity and posture of the neck, and access the typing performance for VR typing by facilitating the representation of virtual hands. We hypothesized that the virtual hand representation would increase typing accuracy and reduce typing injury risk for the hand and shoulder, as well as increase the involvement in typing, which we would verify from neck muscle activity and typing speed, which would be comparable to no VR typing. We also hypothesized that the virtual hand representation would have higher usability for VR typing.

## 2. Methods

### 2.1. Participants

Fifteen participants (8 males and 7 females) were recruited for this study. None of the participants had a history of musculoskeletal diseases in the upper extremity or musculoskeletal pain in the neck and shoulder regions during or within seven days before the study. The Institutional Review Board of the university approved the experimental methodology, and all subjects provided written consent before taking part in the study. The demographic and anthropometric information of participants is summarized in Table 1.

Table 1. Mean and standard deviation of age, height, weight, and typing skill of participants.

|  | Age (years) | Height (cm) | Weight (kg) | Typing Skill (WPM) |
|---|---|---|---|---|
| Average | 25.0 | 168.2 | 72.5 | 29.0 |
| Standard deviation | 3.0 | 9.3 | 14.1 | 7.0 |

Note: WPM = word per minute

## 2.2. Apparatus

Oculus Quest 2 (Figure 1) with 1832x1920 resolution per eye, fast-switch LCD, 360 degrees field view with 6 degrees of freedom head was used for the study. The processor was Qualcomm® Snapdragon™ XR2 Platform and had a 72 - 120 Hz refresh rate. The headset had approximately 503 grams of weight and an integrated hand-tracking system. The headset was comparable to the keyboard and the degree of freedom and high refresh rate were suitable for hand tracking.

A Logitech K830 keyboard was used for the study. It has an area of 36.7 cm x 12.53 cm and is 1.65 cm thick. It was integrated with 2.4 GHz Wireless Technology and Bluetooth smart feature. The maximum operating distance is up to 10 meters. The keyboard is compatible with both VR, android. It can be operated on charged uncabled or directly charging form. The keys glow with red borders while in operation, and every keypress can be detected as the red border changes into blue with every keypress. While using the VR, the virtual hand could be visible for typing if the hand-tracking option of the VR was turned on.

*2.3. Task Conditions*

In randomized order, typing tasks were performed by the 15 participants. Participants were trained with the VR interface and briefed about the terms and conditions of the tasks before conducting the task. To perform the tasks on the VR interface, the participants wore Oculus Quest 2. A Logitech K830 keyboard was provided to pair up with the laptop and VR.

The task was divided into two task conditions. In both conditions, the participants were wearing VR headsets and typing using the conventional keyboard (Figure 1). During the VR with hand representation condition participants were able to see their virtual hand because the VR hand tracking option was on and in the VR without hand representation condition participants were not able to see their virtual hand because the VR hand tracking option was off.

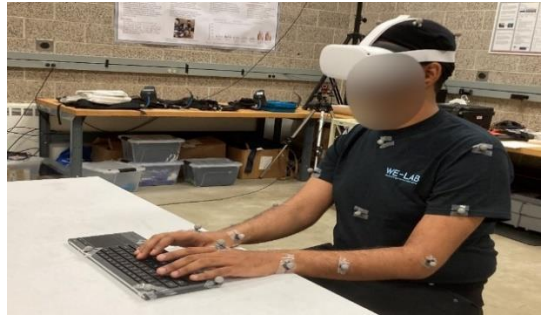

Figure 1. Example of the experimental condition.

*2.4. Measures*

**Performance measure**

The typing test was performed on the TypingTest online platform for a 5-minute medium test for each task condition (Cloud Kayak Labs, Inc, 2022). The typing performance had three parameters: actual speed in word count per minute, accuracy percentage, and the net speed in words per minute.

The actual speed was the count of the word a participant typed within a minute, and the accuracy was the correct percentage of word count within five minutes. The net speed was the multiplication of the actual speed and the decimal value of the accuracy percentage.

**Kinematic data**

An eight-camera optical motion capture system (Flex 13; Optitrack; Natural Point, OR) with reflective markers (14 mm) was used to sample the tasks' upper extremity kinematic data at 120 Hz. Conventional upper extremity with 27 markers were used to collect the motion tracking data.

A digital zero-phase 4th order Butterworth filter with a cutoff frequency of 6 Hz was used to filter the raw kinematic data (Motive; Optitrack; Natural Point, OR). The wrist and shoulder flexion/extension angles, the neck axial rotation angle, the neck lateral flexion angle, and the abduction/adduction angle were computed using biomechanical analysis software (Visual3D; C-Motion Inc., Germantown, MD). The rotation matrix between the anatomical coordinate systems of the head and trunk was used to calculate neck angles. The instantaneous orientations of the anatomical axes in the upper arms and the trunk were used to compute shoulder angles. The wrist angles were calculated using instantaneous orientations of the anatomical axes in the forearms and hands. Using the Amplitude Probability Density Function, these kinematic variables were condensed into 50th-percentile values (JONSSON, 1982).

**Electromyography**

Using wireless surface electrodes (Trigno Avanti Wireless Sensors, Delsys, Natick, MA, USA) and a Samsung Galaxy Tab S5e as a wireless logger, muscle activity in the left and right splenius capitis (LSC, RSC) was recorded (EMG Plots app; Delsys Incorporated; Natick, MA, USA). According to the European Recommendation for Surface Electromyography (EMG), the skin was prepped, the muscles were identified, and the electrodes were placed (Li and Li, 2013). EMG sensors were placed bilaterally at the midpoint of a line connecting the C7 and the mastoid process to collect signals from the splenius capitis muscles. Root mean squared (RMS) data of the mentioned muscles was collected with a 100 ms sampling window.

With a 2-minute interval between sessions, three trials of Maximum Voluntary Contractions (MVCs) were recorded at the end of the experiment for LSC and RSC. The participants attempted to twist their heads backward while sitting in a slightly flexed position as cervical extensor MVC EMG data were recorded (Kang and Shin, 2017). To standardize the EMG data, the 95th percentile peak of the readings from three MVCs was employed (Odell et al., 2007). Using the amplitude probability distribution function (APDF), the normalized muscular activity (percent MVC) was summarized as the 50th percentile (median) (JONSSON, 1982).

**Usability**

A widely used standardized test for evaluating perceived usability is the System Usability Scale (SUS) was used for this study. According to Lewis and Sauro (2009), in industrial usability studies, the SUS accounted for 43% of post-study questionnaire usage (Lewis and Sauro, 2009). Each participant responded to the survey twice times, each time immediately following the completion of the task condition. Conceptually, the first step in scoring was to translate raw item scores into adjusted scores, which are also referred to as "score contributions" and range from 0 (the worst rating) to 4 (the best rating). The adjustments are different for the odd- and even-numbered items, which correspond to the positive and negative-tone items, respectively. 1 was added to the raw score for items with odd numbers, and 5 was added to the raw score for goods with even numbers.

To obtain the standard SUS score, the modified scores were added together, and the result was multiplied by 2.5. Acceptable usability was defined as a SUS score of 70 or higher out of 100. Acceptable usability was further classified into three levels: good (70-80), excellent (80-90), and best possible (90-100).

*2.5. Statistical analysis*

Median values (50th percentile) were summarized for joint angles and muscle activities using the amplitude probability density function (APDF). For this study, the Wilcoxon signed-rank test (a nonparametric test method based on differences for the study of matched-pair data) was conducted, which determined if the differences in the two task conditions were statistically significant. Statistical significance was set as 0.05 (Woolson, 2008). Moreover, descriptive analysis (median and interquartile range) was calculated to summarize all subjects' data for typing performance, joint angles, muscle activity, and usability.

**3. Results**

*3.1 Typing Performance*

Table 2 showed significantly higher typing speed and accuracy for VR with hand typing conditions compared to VR without hand representation (p's < 0.001). Especially, virtual hand representation increased the net speed (WPM) by 171.4% compared to the without virtual hand representation.

Table 2. Median (interquartile range) of typing performance measures with and without virtual hand representation.

| Variables | Without Hand Representation | With Hand Representation | p-value |
| --- | --- | --- | --- |
| Actual Speed (WPM) | 8 | 20 | < 0.001* |
|  | (7) | (8) |  |
| Accuracy (%) | 97 | 98 | < 0.001* |
|  | (5) | (5) |  |
| Net speed (WPM) | 7 | 19 | < 0.001* |
|  | (6) | (8) |  |

Note: * denotes that p-value < 0.01.

*3.2 Joint Angles*

All joint angles of the neck, shoulders, and writs were significantly different with and without the VR hand representation ($p < 0.005$). Especially, the right shoulder abduction was found 6° higher for with VR hand representation than without VR hand representation ($p = 0.003$). Tying with virtual hand representation decreased the right wrist extension angle by 4° but increased the left wrist extension angle by 4° compared to the without virtual hand representation ($p < 0.001$). In addition, typing with virtual hand representation increased the neck flexion angle by 3° relative to the without virtual hand representation ($p = 0.005$).

Table 3. Median (interquartile range) of neck, shoulder and wrist median (50$^{th}$ percentile) joint angles with and without virtual hand representation.

| Joint Angles (°) | Without | With | p-values |
| --- | --- | --- | --- |

|  | Hand Representation | Hand Representation |  |
|---|---|---|---|
| Neck Flexion | 11 | 14 | 0.005* |
|  | (23) | (29) |  |
| Neck Rotation | 4 | 3 | 0.002* |
|  | (14) | (7) |  |
| Left Shoulder Flexion | 49 | 53 | < 0.001* |
|  | (18) | (26) |  |
| Left Shoulder Abduction | 4 | 4 | < 0.001* |
|  | (11) | (7) |  |
| Right Shoulder Flexion | 46 | 45 | < 0.001* |
|  | (14) | (11) |  |
| Right Shoulder Abduction | 1 | 7 | 0.003* |
|  | (9) | (13) |  |
| Left Wrist Extension | 23 | 19 | < 0.001* |
|  | (20) | (18) |  |
| Left Wrist Ulnar Deviation | 28 | 25 | < 0.001* |
|  | (29) | (35) |  |
| Right Wrist Extension | 16 | 14 | < 0.001* |
|  | (4) | (6) |  |
| Right Wrist Ulnar Deviation | 23 | 27 | < 0.001* |
|  | (12) | (11) |  |

Note: * denotes that p-value < 0.01. For the neck rotation angle, the negative value indicates the right rotation.

## 3.3 Muscle Activity

Neck muscle activities were significantly affected by the VR hand representations ($p < 0.001$). When typing with virtual hand representation, the neck muscles' normalized muscular activity (%MVC) was significantly higher (by 21.2% for LSC and 6.1% for RSC) than when typing without virtual hand representation.

Table 4. Median (interquartile range) of neck median ($50^{th}$ percentile) muscle activity with and without virtual hand representation.

| Normalized Muscle Activity (%MVC) | Without Hand Representation | With Hand Representation | p-values |
|---|---|---|---|
| Left Splenius Capitis (LSC) | 13.46 (18.33) | 16.32 (20.91) | < 0.001* |
| Right Splenius Capitis (RSC) | 6.60 (7.77) | 7.00 (9.62) | < 0.001* |

Note: * denotes that p-value < 0.01.

## 3.4 Systems Usability Score (SUS)

The SUS was significantly influenced by the VR hand representation. SUS score for VR without hand representation was decreased by 10 compared with the with virtual hand representation (Table 5).

Table 5. Median (interquartile range) of SUS Score

| Typing Conditions | Without Hand Representation | With Hand Representation | p-value |
|---|---|---|---|
| SUS Score | 40 (30) | 50 (27.5) | < 0.001* |

Note: * denotes that p-value < 0.01.

The summed values of individual question were compared with and without the virtual hand representation (Figure 2). The most pronounced increase of the usability with the virtual hand representation was easiness of use (Q3) followed by frequent use (Q1), and quick learning (Q7).

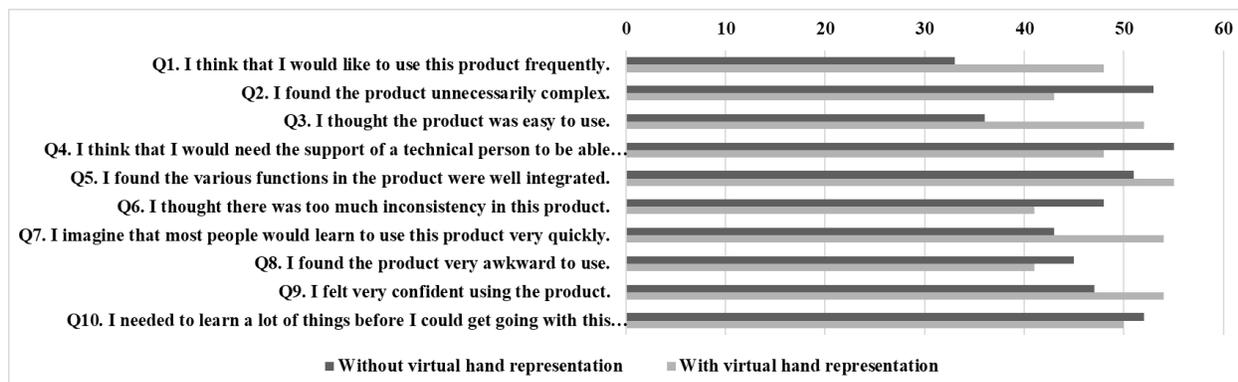

Figure 2. The summed values of each usability question with and without virtual hand representations.

## 4. Discussion

This study aimed to evaluate the effect of VR hand representation on the typing performance, upper extremity joint angles, neck muscle activity, and usability. It was found that the aforementioned measures were significantly affected by the VR hand representation. The virtual hand representation improved typing performance and usability compared to the without virtual hand representation. There were notable changes of the upper extremity angles and variability of the motions with and without the virtual hand representations. Neck muscle activities were increased with the virtual hand representation while typing.

In this study, virtual hand representation significantly improved typing speed and accuracy compared to without hand representation. Especially, improvement of the net speed (171.4% increase) was more notable than the accuracy (1.0% higher). Participants were allowed to correct their typos during typing, and this could lead to more comparable accuracy performance with and without the virtual hand representations. Based on a previous study by Grubert et al. (2018), virtual hand representation could enhance typing performance by providing users with a more realistic interface, which inclined with the result of this study (Grubert et al., 2018b). Moreover, according to Huang et al. (2015), the use of virtual hand representation may have increased the user's involvement in typing, for this study which resulted in improved typing speed and accuracy (Huang et al., 2012).

There was different typing behavior observed with and without the virtual hand representations. Typing with virtual hand representation showed increased neck flexion by 3° compared to without virtual hand representation. This could be related to the presence of the virtual hand and typing skills of participants. Majority of participants' typing skill level was at a beginner level (WPM < 40). They tended to look their virtual hand in VR environment for accurate and efficient typing. This could result in increased neck flexion angle. Another reason could be the limited field of view while using the head-mounted display. Since the filed of view with VR head-mounted display is narrower than actual human eye's filed of view, participants could move their head more to look at their hands and keyboard in VR environment.

For typing without virtual hand representation, there was a notable reduction of the right shoulder abduction by 6° and increase of the left/right wrist extensions by 4°. Since users could not see the locations of their hands in VR environment, their confidence of typing could be decreased. This could result in more restricted typing behavior such as lower angles of the shoulder abduction and higher angles of the wrist extension. In addition, users tended to push the backspace key more often to correct their typos when typing without virtual hand representation. This could naturally increase the wrist extension angles. Long-term exposure of typing without the virtual hand representation could increase the risk of carpal tunnel syndrome, tendinitis, and bursitis (Nag, 2019; Rempel et al., 2008).

Typing without the virtual hand representation revealed more variability of the upper extremity motions compared to with the virtual hand representation. Especially, participants showed increased variability (up to 50%) of the neck rotation while typing without virtual hand representation. Although participants could not see the virtual hand, they were able to see the highlighted keys when they pushed the keys. Participants could navigate the keys by rotating their heads to understand which keys they typed. Frequent head rotations in VR environment could increase VR sickness and neck discomfort (Kim and Shin, 2020).

Bilateral neck muscle activities were significantly increased with virtual hand representation compared to the without hand representation. The increased neck muscle activities could be explained by the increase of the neck flexion angle with virtual hand representation. Increased neck flexion angle is known to be associated with increased neck moment-arm between the center of the mass of the head and the cervical joint (C7 level) (Syamala et al., 2018). To counterbalance the increased gravitational moment due to the neck flexion, neck muscle activities could be increased. This pattern was consistent with the previous study assessing different neck muscle activities by varying vertical target locations during VR environment (Penumudi et al., 2020). Increased neck muscle activities could be also associated with the typing behavior. Typing with virtual hand representation substantially increased the typing speed by 171.4% compared to the without hand representation. This more frequent arm and hand movements could naturally induce more neck muscle activities.

The usability score was significantly higher with virtual hand representation compared to without hand representation. According to the participants, virtual hand representations were more realistic and provided a more natural way to interact with the VR environment. They scored high values of easiness of use, willing to use frequently, and learning quickly of the product. According to previous research, virtual hand representation increases the sense of presence in virtual environments (Grubert et al., 2018b). However, based on the SUS standard, both conditions did not meet acceptable usability conditions (SUS > 70). This suggests that continuous improvement of virtual hand representations could be needed to increase the usability level. Different skin tone, shape, size, and joint visibility could be important design factors to increase the visibility and usability while typing.

Although this study was carefully designed and controlled, this study had some limitations. First, this study did not consider long-term exposure of typing and thus did not encounter and evaluate the effect of a learning curve and the physical fatigue from prolonged exposure to VR typing. Second, the participants had mostly lower levels of typing expertise, and their expertise level was very comparable each other. Hence, further study would be needed to identify and distinguish the effect of virtual hand representations among different expertise levels of typists. Lastly, the experiment was also carried out in laboratory settings with the participation of university students. Future studies could investigate the efficacy of virtual hand representation in real office environments setting.

## 5. Conclusion

A virtual hand representation in VR typing could improve the typing performance and usability compared to the without virtual hand representations. Without virtual hand representation, users tended to have increased wrist extension and reduced right shoulder abduction. Variability of the neck rotation was substantially increased as well. This alternated typing behavior without the virtual hand representation could increase the risk of the musculoskeletal disorders and VR sickness. Although usability was significantly increased with the virtual hand representation, SUS scores still did not meet the acceptable usability level. This indicates that continuous improvement of the virtual hand design could be needed by considering skin tone, size, and anatomical structures. Future study could investigate the effects of virtual hand representation among participants who have various typing skill sets from beginner to the expertise.